\documentclass{article}
\usepackage{PRIMEarxiv}
\usepackage{array}
\usepackage{multirow}
\usepackage[utf8]{inputenc} 
\usepackage[T1]{fontenc}    
\usepackage{hyperref}       
\usepackage{url}            
\usepackage{booktabs}       
\usepackage{amsfonts}       
\usepackage{nicefrac}       
\usepackage{microtype}      
\usepackage{lipsum}
\usepackage{fancyhdr}       
\usepackage{graphicx}       
\graphicspath{{media/}}     
\usepackage{mdframed}
\usepackage{caption}
\usepackage{graphicx}
\usepackage{longtable}

\title{Enhancing Anomaly Detection in Financial Markets with an LLM-based Multi-Agent Framework

}

\author{
  Taejin Park
  \thanks{\textit{The views expressed here are those of the author only, and not necessarily those of the BIS.}
  \textbf{}} \\
  Bank for International Settlements (BIS) \\
  Basel, Switzerland \\
  \texttt{taejin.park@bis.org} \\
  }
\let\oldmaketitle\maketitle
\renewcommand{\maketitle}{\oldmaketitle\setcounter{footnote}{0}}
\begin{document}
\maketitle

\begin{abstract}
This paper introduces a Large Language Model (LLM)-based multi-agent framework designed to enhance anomaly detection within financial market data, tackling the longstanding challenge of manually verifying system-generated anomaly alerts. The framework harnesses a collaborative network of AI agents, each specialised in distinct functions including data conversion, expert analysis via web research, institutional knowledge utilization or cross-checking and report consolidation and management roles. By coordinating these agents towards a common objective, the framework provides a comprehensive and automated approach for validating and interpreting financial data anomalies. I analyse the S\&P 500 index to demonstrate the framework's proficiency in enhancing the efficiency, accuracy and reduction of human intervention in financial market monitoring. The integration of AI's autonomous functionalities with established analytical methods not only underscores the framework's effectiveness in anomaly detection but also signals its broader applicability in supporting financial market monitoring.
\end{abstract}


\section{Introduction}
Anomaly detection plays a central role in monitoring financial markets, serving as the foundation upon which analysts and statisticians build their understanding of market events and safeguard integrity. This process involves identifying data points that deviate significantly from standard patterns, signalling potential irregularities for further investigation. Despite the wealth of established quantitative techniques for identifying anomalies, a fundamental challenge persists in setting appropriate thresholds that effectively flag data points for further examination. This challenge is underscored by the significant need to maintain a delicate balance between Type 1 errors (false positives) and Type 2 errors (false negatives)—a balance that is essential for distinguishing genuine market anomalies from noises. Setting the threshold too low may result in an overload of false alarms, potentially obscuring truly significant anomalies, whereas a threshold set too high risks missing early warning signs.\par
Anomaly detection methods have conventionally relied heavily on pre-defined quantitative algorithms. Recent advances, however, have seen a shift towards incorporating deep learning-based algorithms to enhance detection capabilities (Chalapathy and Chawla, 2019\cite{chalapathy2019deep}). Despite these technological advances, the core challenge of anomaly detection persists. Upon the identification of potential anomalies, system-generated alerts necessitate users to initiate a comprehensive verification process. This predominantly manual procedure involves a variety of complex actions that are challenging to automate, largely because of the qualitative aspects of the assessment. In such contexts, human expertise and judgment play key roles. The nuanced understanding and experience of professionals in interpreting data and comprehending market contexts are indispensable for the accurate validation and interpretation of results.\par
The emergence of Large Language Models (LLMs) and their integration into autonomous agents opens new opportunities for validation tasks within financial market data analysis. Recent studies, including those by Wang et al. (2023)\cite{wang2023survey}, underscore the emerging capabilities of LLM-based agents in assuming roles traditionally reliant on manual human intervention. The inherent autonomous nature of these agents renders them particularly apt for applications requiring rapid, scalable and nuanced data analysis. \par
Another notable advance in LLMs is the development of a multi-agent framework. Park et al. (2023)\cite{park2023generative} suggest a novel mode of interaction among LLMs that mimics human collaborative dynamics. This framework allows individual LLMs to specialise in distinct areas of expertise, enabling them to work in concert towards a common goal. The synergy achieved through this collaboration enhances the overall performance of the LLM ecosystem, as evidenced by the specialised and collaborative efforts detailed by Li et al. (2024)\cite{li2024agents}. Moreover, the scope of application for the multi-agent framework transcends routine tasks. Boiko et al. (2023)\cite{boiko2023emergent} showcase its aptitude in conducting complex scientific research. This capability indicates that when deployed within such a framework, LLMs can effectively support intricate and knowledge-intensive tasks.\par
These recent technology advances offer a pathway to significantly streamline, and potentially automate, the labour-intensive processes of traditional financial market data analysis. This paper introduces a framework designed to replicate and enhance the financial market data validation workflow. By employing a multi-agent AI model, the framework intends to harness the potential of AI to elevate efficiency while maintaining, possibly augmenting, the rigor and thoroughness of established data analysis methodologies. The overarching goal of this initiative is to merge AI's autonomy with the traditional analysis methods, which can redefine the paradigm of data analysis in financial markets.\par

\section{Proposed Structure of Multi-Agent AI Framework for Anomaly Detection and Analysis}
\begin{figure}[ht!]
    \centering
    \includegraphics[width=1\linewidth]{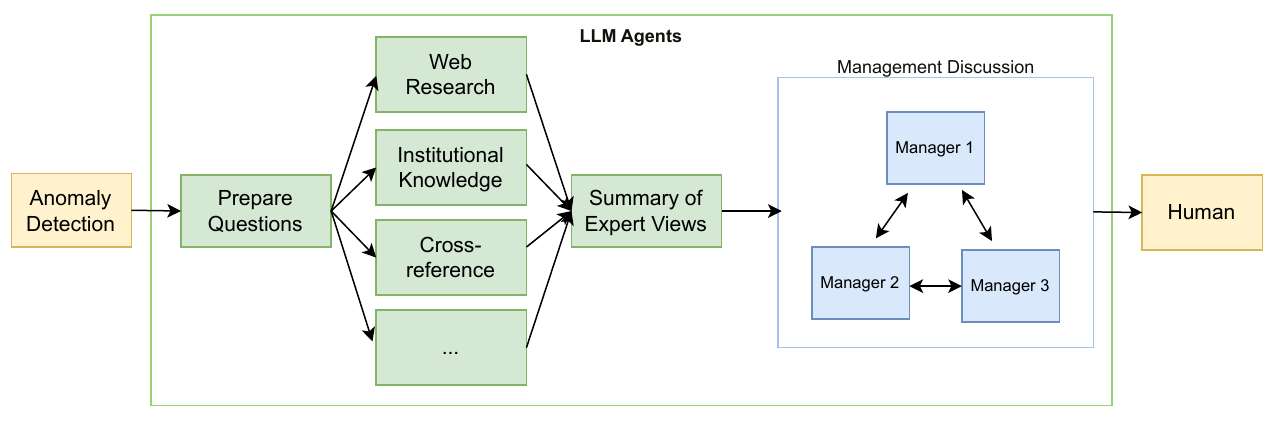}
    \caption{A multi-agent framework of validation process for data anomalies}
    \label{fig:fig1}
\end{figure}

The proposed framework, as depicted in Figure \ref{fig:fig1}, illustrates the advanced methodology for anomaly detection within financial markets, leveraging both established statistical techniques and the application of LLM-based multi-agents. The workflow initiates with identifying anomalies in tabular financial data through existing detection methods, ranging from basic techniques like rule-based methods and z-scores to more sophisticated approaches including unsupervised clustering and deep learning-based methods. \par
Once anomalies are detected, the data is introduced into the LLM-based multi-agent framework, designed to enhance the validation and interpretation of these anomalies. The framework's operational sequence is as follows: \par
\begin{enumerate}
  \item \textbf{Data Conversion by Initial Agent}: The first agent specialises in converting tabular data into a format and structure suitable for LLM processing. For instance, this agent is responsible for formulating questions in a way that allows subsequent agents to understand the identified issues and effectively utilise their LLM engines to process the information. In this phase, the role of extensive metadata is central, as it aids in the proper formulation of questions in a comprehensible format for LLMs.
  \item \textbf{Specialised Data Expert Agents}: The questions are then addressed to a group of expert agents, each with a distinct specialisation:
  \begin{itemize}
    \item \textbf{Web Research Agent}: This agent verifies the authenticity of anomalies by researching web-based resources, such as press releases from data publishers, major news articles or social media posts.
    \item \textbf{Institutional Knowledge Agent}: Functioning as an experienced market analyst, this agent leverages an extensive domain knowledge to provide context and explanations for the detected anomalies. Knowledge of this agent can be derived from past analyses, correspondence with data providers, internal documentation about previous issues, statistical methodologies, etc.
    \item \textbf{Cross-Checking Agent}: Dedicated to validating data through cross-referencing with other reliable sources, this agent plays an essential role in confirming or disputing the anomalies identified. Even when identical datasets are not available, the agent can have the capability to consult analogous data series that usually show similar trends. For instance, it can compare different stock indices that represent the same market or examine government bond yields of similar maturities.
    \item The framework is designed to be adaptable, allowing for the integration of additional agents based on the specific nature of the data in question. This flexibility ensures that the system can be tailored to address unique characteristics and requirements of different data sets, enhancing its effectiveness in handling a wide range of scenarios.
  \end{itemize}
  \item \textbf{Consolidation and Reporting Agent}: An agent specialised in report synthesis consolidates the insights from all data expert agents, crafting a summary report that highlights the key findings. 
  \item \textbf{Management Discussion}: Upon the consolidation of expert analyses into a summary report, this document is forwarded to a panel of management agents. These agents, each focusing on distinct areas, engage in a review and discussion of the report's findings. Mirroring real-world organizational dynamics, these management agents are engineered to adopt high-level perspectives, contrasting with the detail-oriented focus of the data expert agents. This design ensures that strategic insights and broader contexts are considered in the decision-making process. Through their deliberations, the management agents exchange views, debate interpretations, and evaluate the implications of the findings. At the end of the discussion, these agents reach a conclusion on the next course of action.
  \item \textbf{Report to Human}: The consensus reached by the management agents on the recommended course of action is then communicated to a human analyst. This step represents a critical interface between the AI-driven analysis process and human decision-making. The human analyst is equipped with the synthesized information, encompassing both the nuanced details discovered by the expert agents and the strategic recommendations by the management agents. This comprehensive briefing enables the human analyst to make well-informed decisions regarding further investigation, the implementation of corrective measures, or any other necessary actions in response to the identified anomalies. The final decision-making step rests with the human analyst who can leverage the depth and breadth of insights generated by the AI.
\end{enumerate}
\par

This proposed multi-agent AI framework offers a comprehensive solution that automates the process of anomaly detection in tabular data, follow-up analysis and reporting. This workflow can not only improve efficiency but also enhance the accuracy and reliability of financial market analysis. By reducing the reliance on manual processes, the framework presents the potential to reduce human error and bias. Furthermore, the rapid processing capabilities of the AI agents could shorten the time from anomaly detection to action, enabling more timely and effective responses to market anomalies.\par

\section{Demonstration of the Multi-Agent AI Framework Using S\&P 500 Series}
This section offers a hands-on demonstration of the multi-agent AI framework, applied to financial market data—specifically, a daily series of the S\&P 500 index spanning from 1980 to 2023. This example explains how the LLM-powered multi-agent model processes and analyses real-world financial data, illustrating each phase from anomaly detection to the concluding decision-making process. By utilising the well-known S\&P 500 series as a test case, I aim to underscore the framework's proficiency in navigating the intricacies of financial datasets. The examples provided in this section are actual outcomes from the fully automated, custom-developed framework.\par

\subsection{Outlier Detection}
\begin{figure}[ht!]
    \centering
    \includegraphics[width=1\linewidth]{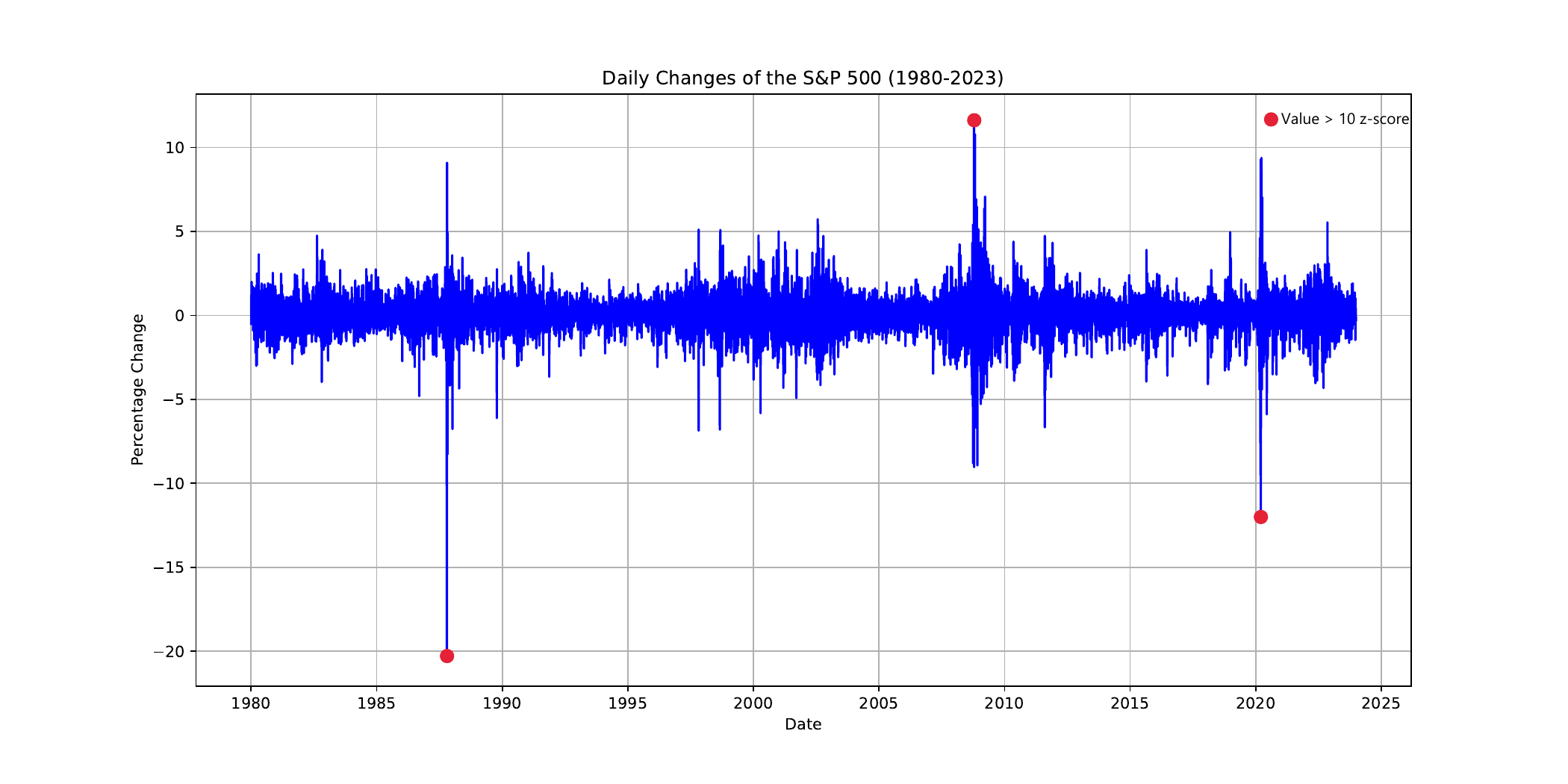}
    \caption{Anomalies identified in the S\&P500 series}
    \caption*{Source: Bloomberg.}
    \label{fig:fig2}
\end{figure}

The initial phase of this demonstration involves outlier detection, executed by applying the z-score method to the daily percentage changes observed in the S\&P 500 series. A deliberately high threshold of 10 z-scores is chosen to pinpoint significant outliers, ensuring a focus on the most pronounced deviations. As a result, three outliers are identified on 19 October 1987, 13 October 2008 and 16 March 2020 (Figure \ref{fig:fig2}). Additionally, to challenge the framework’s discernment capabilities, three missing values are deliberately inserted into the dataset. This approach is designed not only to assess the framework’s capacity for identifying substantial anomalies but also to evaluate its capacity in differentiating between authentic outliers and intentionally introduced inaccuracies. The careful introduction of both legitimate outliers and potential errors crafts a nuanced testing environment, allowing for a thorough assessment of the multi-agent AI model’s adeptness in managing the complexities inherent in real-world financial data. Following detection, these data points are converted into a format suitable for machine processing, as illustrated in Table \ref{table:1}, setting the stage for the following analysis by the AI agents.\par
In the preparation of input data, integrating both tabular data and corresponding metadata is critical for the efficacy of the AI system. Access to metadata—encompassing details such as the data's name, source, frequency, description and coverage—is necessary for the AI to fully understand and contextualise detected anomalies. This integration enables the AI system to interpret tabular data with greater accuracy and to maximize the utilisation of the knowledge acquired during pre-training processes of LLMs.\par

\begin{table}[h]
\centering
\caption{Data for validation with metadata}
\begin{tabular}{|c|p{5.5cm}|p{5.6cm}|} 
\hline
& \centering Original data format (Python) & \centering Machine-readable format \cr
\hline
Data & \texttt{BB:D:SPX\_INDEX:PX\_LAST} \texttt{1987-10-19\hspace{1.1cm}-20.46692607 1987-10-20\hspace{1.1cm}NaN 2008-10-13\hspace{1.1cm}11.5800360312 2008-10-14\hspace{1.1cm}NaN 2020-03-12\hspace{1.1cm}NaN 2020-03-16\hspace{1.1cm}-11.984050283} & \texttt{\hspace{-0.1cm}\raggedright\{"BB:D:SPX\_INDEX:PX\_LAST": \{"1987-10-19":-20.46692607 "1987-10-20":null, "2008-10-13":11.5800360312, "2008-10-14":null, "2020-03-12":null, "2020-03-16":-11.9840502837\}\}} \\ \hline
Metadata & \multicolumn{2}{p{11cm}|}{\raggedright\{'original\_frequency\_code': 'Day', 'CURRENCY': 'USD', 'DATA\_DESCR': 'The S\&P 500 is widely regarded as the best single gauge of large-cap U.S. equities and serves as the foundation for a wide range of investment products. The index includes 500 leading companies and captures approximately 80\% coverage of available market capitalization.', 'PRICING\_SOURCE': "Standard \& Poor's", 'REF\_AREA': 'US', 'TITLE': 'S\&P 500 INDEX'\}} \\ \hline
\end{tabular}
\label{table:1}
\end{table}

\subsection{Agent for Formulating Data Questions}
Upon receiving outlier data and associated metadata, the agent tasked with formulating data questions plays a key role in the initial stage of anomaly validation. This agent's output comprises questions designed to probe the validity and context of the identified outliers. The questions formulated by this agent serve multiple purposes: they aim to confirm the nature of the anomalies detected, understand their significance within a historical and market context and prepare relevant LLM-suitable questions for further verification. Table \ref{table:2} demonstrates how the agent is directed and its responses in addressing the outlier events within the S\&P 500 Index.\par
The output generated by the agent reflects a human-like response, indicative of the satisfactory integration of tabular data and LLMs. This sophistication is evident in several key aspects:

\begin{itemize}
    \item \textbf{Contextual Awareness}: Despite the absence of explicit event information in the provided data or metadata, the agent infers and incorporates relevant historical contexts, such as Black Monday and the impact of COVID-19 on financial markets. This ability to relate numerical outliers to significant real-world events demonstrates the agent's contextual understanding and its use of pre-trained data to enrich analysis.
    \item \textbf{Adaptability}: The agent's questions not only seek to validate the nature and accuracy of the outlier data but also intelligently speculate on possible explanations for these anomalies, such as suggesting whether a data point represents a percentage drop, a point drop, or another measurement. This adaptability ensures a comprehensive validation process that accounts for various possible causes of the problem.
    \item \textbf{Efficiency}: By grouping similar questions and summarising inquiries related to missing values into a single question, the agent efficiently manages the context window. This approach optimises the interaction with subsequent LLM-based analysis stages, ensuring that the questions remain within acceptable limits for processing and analysis. This efficiency is crucial for maintaining the system's performance, scalability and responsiveness.
    \item \textbf{Pre-trained Knowledge Utilization}: The agent's ability to add additional information and offer possible explanations based on pre-trained knowledge highlights the powerful integration of LLMs into the framework. This integration allows the system to leverage vast amounts of historical data and insights, enhancing the depth and accuracy of the anomaly validation process.
\end{itemize}

\begin{longtable}{|p{\textwidth}|}
\captionsetup{width=.75\textwidth}

\caption{Instruction to and output of the Agent for formulating data questions\label{table:2}} \\
\hline
\textbf{INSTRUCTION TO AGENT}:\\

You are a critical data expert who will generate questions based on given data in JSON format. All of values in this table are outliers that require review. You will also be given metadata to understand what's in the data.\\
Please prepare questions for each data column to validate whether these outlier values are correct or not. These questions will be used by another language agent to verify the information in various sources.\\
If there's no outlier, no question is needed.\\
\\
\textbf{OUTPUT FROM AGENT}:\\

Based on the provided outlier data and metadata, here are questions to validate the outlier values:\\
\begin{enumerate}
    \item 1987-10-19: -20.47
    
    Can you confirm if the S\&P 500 Index had an unusual decrease on October 19, 1987, and if the value of -20.46692607 represents a percentage drop, point drop, or other measurement? Was this value part of the event known as Black Monday?
    \item 2008-10-13: 11.58
    
     Can you verify if the S\&P 500 Index experienced a significant increase on October 13, 2008, and if the value of 11.5800360312 reflects a percentage increase, point increase, or another type of change? Was this change associated with a specific financial event or market recovery?
    \item 2020-03-16: -11.98
    
    Could you check if there was a substantial drop in the S\&P 500 Index on March 16, 2020, and whether the value of -11.9840502837 indicates a percentage drop, point drop, or a different measure? Was this drop connected to the market's response to COVID-19 developments?

\setlength{\parindent}{0cm}Additionally, the data for 1987-10-20, 2008-10-14, and 2020-03-12 are missing (null). Here are questions to address this:
    \item Were there no recorded changes in the S\&P 500 Index values on October 20, 1987; October 14, 2008; and March 12, 2020, or are the null values indicating missing data for these dates? If the data is missing, is there a known reason for the absence of data on those specific dates?
\end{enumerate}

Overall, the questions seek to confirm whether the outlier values correctly reflect the historical performance of the S\&P 500 Index on the given dates and to understand the reason behind any null values. \\
\hline
\end{longtable}

\subsection{Data Expert 1: Web Research}
Upon receiving the formulated questions, the first data expert agent employs a web research platform, such as Tavily\footnote{For details, see https://tavily.com.}, that integrates LLM capabilities with access to real-time web information. This integration allows the agent to conduct investigations into each question, providing a reliable verification process based on current and historical web data. Table \ref{table:3} shows the instruction to this agent and its output, detailing its findings and insights.\par
The outcomes of the web research expert agent's analysis offer factual validations and insights into the nature of the detected anomalies. By associating these anomalies with significant historical or contemporary events, the agent not only corroborates the accuracy of the outlier data with reliable sources but also enhances the analytical context. Furthermore, it accurately suggests the appropriate unit of measurement for the values, specifically in percentages. The designation of 'Uncertain' classification for missing data points highlights the inherent challenges in validating such data, particularly null values or seemingly insignificant events. This underscores the reality that major news media or press releases may not always provide relevant information for every identified anomaly, posing a challenge in ensuring the completeness and accuracy of the data analysis by a single expert agent.

\begin{longtable}{|p{\textwidth}|}
\captionsetup{width=.75\textwidth}

\caption{Instruction to and output of the Agent for web research\label{table:3}} \\
\hline
\textbf{INSTRUCTION TO AGENT}:\\

You are a web searching expert to verify some data-related questions. When I give you questions, can you search web and verify the information? As you need to verify all the bullets, please respond concisely. Your answer starts with Correct, Incorrect, or Uncertain. Then provide the reasoning in one sentence.\\
\\
\textbf{OUTPUT FROM AGENT}:
\begin{enumerate}
    \item \textbf{Correct}: The S\&P 500 Index had an unusual decrease on October 19, 1987, which represented a percentage drop, and this was part of the event known as Black Monday.
    
    - The Federal Reserve history mentions the Dow Jones Industrial Average dropped 22.6 percent, indicating a similar percentage drop for the S\&P 500 on Black Monday.
    \item \textbf{Correct}: The S\&P 500 Index experienced a significant increase on October 13, 2008, and the value of 11.5800360312 reflects a percentage increase associated with a market recovery during the financial crisis.
    
     - A CNN report from the time confirms an increase of 11.8\%, the index's second-best daily percentage gain at that point, following initiatives to tackle the financial crisis.
    \item \textbf{Correct}: There was a substantial drop in the S\&P 500 Index on March 16, 2020, and the value -11.9840502837 indicates a percentage drop connected to market response to COVID-19 developments.
    
    - CNBC reported a 12\% single-day drop around that time, aligning with the given value as a percentage drop due to COVID-19 concerns.
    \item \textbf{Uncertain}: Without explicit confirmations for these specific dates, it's unclear whether there were no recorded changes or if the null values represent missing data.
    
    - The searches did not provide immediate information regarding the absence of data for the dates in question.
\end{enumerate}
\\
\hline
\end{longtable}

\subsection{Data Expert 2: Institutional (Domain) Knowledge}
This data expert agent analyses data queries using a targeted, predefined knowledge base that focuses on general stock market trends. For illustrative purposes, I have developed a concise knowledge base that details significant financial crisis periods and the fundamental characteristics of stock market indices, for example, high volatilities during financial crises. The outcomes presented in Table \ref{table:4} reveal the agent's capacity to correlate the detected anomalies with its knowledge base, providing insights and rationales derived solely from this curated repository. In this instance, the data series in question is well-known, as are the related historical events and moving patterns, leading to similar responses as those from the previous data expert agent that conducted web research. Nevertheless, the value of this agent is expected to increase when applied to derived datasets or those not widely recognised by the public, where its specialised knowledge base can provide unique and valuable insights.\par
\begin{longtable}{|p{\textwidth}|}
\captionsetup{width=.75\textwidth}

\caption{Instruction to and output of the Agent for institutional knowledge\label{table:4}} \\
\hline
\textbf{INSTRUCTION TO AGENT}:\\

You are a stock market expert to verify some data-related questions. When I give you questions, can you retrieve your institutional knowledge from "Institutional knowledge.docx" and use it for your research to answer the questions. Use web search to verify facts as needed. As you need to verify all the questions, please respond concisely. Your answer starts with Correct, Incorrect, or Uncertain. Then provide the reasoning in one sentence.\\
\\
\textbf{OUTPUT FROM AGENT}:
\begin{enumerate}
    \item \textbf{Correct}: The S\&P 500 Index had an unusual decrease on October 19, 1987, and the value of -20.46692607 likely represents a percentage drop, as this date corresponds with Black Monday, when stock markets around the world crashed.
    \item \textbf{Correct}: On October 13, 2008, the S\&P 500 Index experienced a significant increase, and the value of 11.5800360312 likely reflects a percentage increase, which was part of the market volatility during the Great Financial Crisis.
    \item \textbf{Correct}: The S\&P 500 Index saw a substantial drop on March 16, 2020, and the value of -11.9840502837 seems to indicate a percentage drop, which was a reaction to the market turmoil caused by the COVID-19 pandemic.
    \item \textbf{Uncertain}: The null values for October 20, 1987; October 14, 2008; and March 12, 2020, could indicate missing data; however, without information about trading suspensions, technical issues, or reporting errors, it's unclear why these specific dates have no recorded changes.
\end{enumerate}
\\
\hline
\end{longtable}
\subsection{Data Expert 3: Cross-Checking}
The cross-checking agent conducts a comparison of the data in question with another reliable data source. The agent initiates its verification process by identifying the specific identifier of S\&P 500 Index through the use of an LLM.\footnote{The prompt for this task is generated in Python with the following code line: f"What is the yahoo finance ticker for \{meta\_data['TITLE']\}? Please just return the ticker."}  Upon successful identification, the agent is programmed to access Yahoo Finance, a comprehensive source of financial market data, to fetch corresponding historical data for the dates in question via an Application Programming Interface (API).\par
By matching the data against an external source, the agent confirms the accuracy of the anomalies detected within the dataset. More importantly, as detailed in Table \ref{table:5}, the agent's investigation finally reveals that the missing data points, flagged as null values, are in fact errors rather than intentional omissions or instances of no change. Due to the prominence and the widespread usage of the S\&P 500 Index data, the cross-checking agent's role for cross-checking is particularly impactful in this context. 
\begin{longtable}{|p{\textwidth}|}
\captionsetup{width=.75\textwidth}

\caption{Instruction to and output of the Agent for cross-checking\label{table:5}} \\
\hline
\textbf{INSTRUCTION TO AGENT}:\\

You are a financial market data expert who is specialised in cross-checking data. When I give you some data-related questions, can you answer them by cross-checking with data I provide in JSON? As you need to verify all the bullets, please respond concisely. Your answer starts with Correct, Incorrect, or Uncertain. Then provide the reasoning in one sentence.\\
\\
\textbf{OUTPUT FROM AGENT}:\\
Correct, the provided data confirms the following:
\begin{enumerate}
    \item The S\&P 500 Index had a decrease of approximately -20.47\% on October 19, 1987, which is consistent with the provided percentage and the event known as Black Monday.
    \item The index experienced an increase of approximately 11.58\% on October 13, 2008, aligning with the given percentage, which could be related to a market recovery or specific financial event.
    \item There was a drop of approximately -11.98\% in the S\&P 500 Index on March 16, 2020, which matches the provided percentage and likely relates to the market's response to COVID-19 developments.
    \item The data for October 20, 1987; October 14, 2008; and March 12, 2020, are not null. In fact, the Index values did change on those dates, suggesting the absence of null values and indicating there might be some miscommunication regarding missing data. The values on those dates are 236.83, 998.01, and 2711.02 respectively.
\end{enumerate}
\\
\hline
\end{longtable}

\subsection{Agent for Summary Reports}
This agent undertakes the task of aggregating the assessments from the preceding data expert agents. By combining these insights, the agent ensures that the summary report provides a holistic view of the anomaly validation process, highlighting the accuracy of the identified outliers, correcting any errors and emphasising findings.\par
The summary report serves as a tool for management agents or human analysts, providing them with a distilled overview of the assessment outcomes (Table \ref{table:6}). By presenting validated findings and highlighting any areas of uncertainty or required further investigation, the report facilitates informed decision-making, enabling strategic responses to the identified market anomalies.\par
\begin{longtable}{|p{\textwidth}|}
\captionsetup{width=.75\textwidth}

\caption{Instruction to and output of the Agent for summary reports\label{table:6}} \\
\hline
\textbf{INSTRUCTION TO AGENT}:\\

Your role is to summarise reports from several experts to present to management.\\
Each expert tried to answer data-related questions provided below.\\
Please concisely summarise the experts' answers.\\
\{Initial questions\}\\
\{Experts’ answers\}\\

\\
\textbf{OUTPUT FROM AGENT}:\\
The experts provided consistent insights on the historical performance and outlier values of the S\&P 500 Index for the dates in question. Here's a summary of their findings:
\begin{enumerate}
    \item \textbf{October 19, 1987}: All experts confirm that the S\&P 500 Index had an exceptional decrease on this date, with the value of -20.46692607 representing a percentage drop indicative of the market crash known as Black Monday.
    \item \textbf{October 13, 2008}: The S\&P 500 Index registered a notable increase on this date, which the experts agree reflects a percentage increase of 11.5800360312. This surge was tied to the market recovery efforts during the financial crisis.
    \item \textbf{March 16, 2020}: The experts corroborate that there was a significant drop in the S\&P 500 Index, with the value of -11.9840502837 indicating a percentage decrease associated with the market's reaction to COVID-19 developments.
    \item \textbf{October 20, 1987; October 14, 2008; and March 12, 2020}: While initially uncertain on whether the null values indicated no recorded changes or missing data, the cross-checking expert clarifies that the values for these specific dates were indeed available, reporting Index values of 236.83, 998.01, and 2711.02 respectively, which negates the assumption of missing data.
\end{enumerate}

\\
\hline
\end{longtable}
\subsection{Management Discussion}
The summary report is then presented for management discussion through AutoGen (Wu et al, 2023\cite{wu2023autogen}), facilitating a dialogue among multiple agents aimed at achieving a pre-defined objective – reaching an agreement on the contents of the summary report. At this stage, I define the role of each manager agent to embody broad expertise in areas such as financial markets, macroeconomics and statistical analysis. Then, they review the report and exchange their strategic perspectives autonomously. A moderator agent, specifically appointed for the task, facilitates the discussion, while a designated critic scrutinises the proceedings with a critical eye to maintain the discussion's robustness and comprehensiveness.\par
The objective of this phase is to provide insights from a managerial standpoint, facilitating an understanding of the data through the lens of experienced domain specialists. This approach seeks to expand viewpoints and enrich the analysis prior to the human review phase, thereby ensuring a comprehensive and multi-faceted evaluation of the information at hand.\par
As illustrated in Table \ref{table:7}, the management agents engage in human-like dialogue, contributing additional insights to the summary report based on their areas of expertise. Following the discussion, the management agents collectively agree to endorse the summary report, making it ready for a final review by a human analyst.\par

\begin{longtable}{|p{\textwidth}|}
\captionsetup{width=.75\textwidth}

\caption{Transcript of management discussion on the summary report\label{table:7}} \\
\hline
Financial Market Economist:\\
The summary report provides a clear and concise overview of the historical performance of the S\&P 500 Index on specific dates that were of interest due to their significant market movements. The findings from the financial market data experts are consistent with well-documented historical events and provide valuable insights into the behavior of the S\&P 500 during times of market stress.
\begin{enumerate}
    \item The confirmation of the S\&P 500 Index's unusual decrease on October 19, 1987, aligns with historical records of Black Monday, when stock markets around the world crashed. The value of -20.46692607 as a percentage drop is accurate and represents one of the largest one-day percentage declines in the history of the S\&P 500. This event had far-reaching consequences for financial markets and led to the implementation of new trading rules and systems designed to prevent such crashes in the future.
    \item The significant increase on October 13, 2008, is also well-documented. The value of 11.5800360312 as a percentage increase reflects the market's response to coordinated efforts by global financial authorities to address the liquidity crisis during the 2008 financial crisis. This event is a prime example of how policy interventions can have immediate and substantial effects on financial markets.
    \item The substantial drop on March 16, 2020, is correctly attributed to the market's response to the unfolding COVID-19 pandemic. The value of -11.9840502837 as a percentage decrease is indicative of the high levels of volatility and uncertainty that characterized the market during the early stages of the pandemic. This event highlights the sensitivity of financial markets to global health crises and the rapid impact such events can have on investor sentiment and market stability.
    \item The clarification regarding the null values for October 20, 1987; October 14, 2008; and March 12, 2020, is important. It confirms that there were indeed recorded changes on these dates and that the data is not missing. This information is crucial for maintaining the integrity of historical market data and ensuring that analyses based on this data are accurate.
\end{enumerate}
Overall, the summary report is well-constructed and provides a factual recount of the S\&P 500 Index's performance on the dates in question. It underscores the importance of historical market data in understanding the dynamics of financial markets and the potential impact of external events on market performance. As a financial market expert manager, such data is invaluable for developing strategies, managing risk, and making informed investment decisions.\par
----------------------------------------------------------------------------------------------------------------------------------------------
Macro Economist:\\
As a macroeconomic expert manager, the summary report provides a succinct and accurate reflection of significant macroeconomic episodes as evidenced by the S\&P 500 Index movements on the dates in question. The report's findings are consistent with the historical context of each event and offer a clear understanding of the market dynamics during those periods.
\begin{enumerate}
    \item The October 19, 1987, event, known as Black Monday, is a critical episode in financial history, representing a systemic shock that led to a reevaluation of risk management and regulatory practices. The -20.46692607 percentage drop is a stark reminder of the vulnerabilities in financial markets and the potential for rapid contagion across global markets. This event has been extensively studied in macroeconomics for its implications on market efficiency, investor behavior, and the role of automated trading systems.
    \item The increase on October 13, 2008, reflects the market's response to policy measures during the global financial crisis. The 11.5800360312 percentage increase underscores the effectiveness of coordinated policy interventions in stabilizing financial markets. This episode is significant for macroeconomic analysis as it highlights the interplay between financial stability and monetary policy, as well as the importance of confidence in financial institutions.
    \item The drop on March 16, 2020, due to COVID-19, illustrates the sensitivity of financial markets to exogenous shocks and the rapid reassessment of risk that can occur in response to unprecedented events. The -11.9840502837 percentage decrease is a clear example of how health crises can translate into economic and financial crises, affecting asset prices, consumer behavior, and economic policy.
    \item The clarification on the null values for the dates following each event is crucial for maintaining data integrity. It ensures that any analysis or modeling based on this data is grounded in complete and accurate information. The presence of data for October 20, 1987; October 14, 2008; and March 12, 2020, allows for a more nuanced understanding of market reactions and recovery processes following shocks.
\end{enumerate}
In conclusion, the summary report effectively captures the essence of these macroeconomic episodes and their manifestations in the S\&P 500 Index. It reinforces the importance of historical analysis in understanding market behavior and preparing for future episodes of financial instability. As a macroeconomic expert, such insights are invaluable for policy recommendations, risk assessment, and the development of economic resilience strategies.\par
----------------------------------------------------------------------------------------------------------------------------------------------
Statistician:\\
As a statistics expert manager, my focus is on the accuracy and reliability of the data presented in the summary report, as well as the statistical significance of the market movements on the dates in question.        
\begin{enumerate}
    \item The report correctly identifies October 19, 1987, as the date of an unusual decrease in the S\&P 500 Index, known as Black Monday. The value of -20.46692607 is indeed a percentage drop, which is statistically significant given the historical average daily movements of the index. This event is a statistical outlier and represents an extreme deviation from typical market behavior.
    \item The significant increase on October 13, 2008, is also statistically notable. The percentage increase of 11.5800360312 is well beyond normal market fluctuations and indicates a strong market rebound, likely influenced by external interventions or news. The statistical significance of this movement can be assessed through comparison with the distribution of historical daily returns.
    \item The drop on March 16, 2020, is another example of a statistically significant market movement. The percentage decrease of -11.9840502837 reflects the market's reaction to the uncertainty and fear surrounding the COVID-19 pandemic. This event would be considered an outlier in a statistical model of market returns, highlighting the impact of extraordinary global events on financial markets.
    \item The clarification regarding the null values for October 20, 1987; October 14, 2008; and March 12, 2020, is important from a data completeness perspective. Ensuring that the data for these dates is not missing allows for accurate statistical analysis. Missing data can lead to biased estimates and incorrect conclusions, so confirming the presence of these values is crucial for any subsequent statistical modeling or analysis.
\end{enumerate}
From a statistical standpoint, the summary report provides a clear account of the extreme market movements on the specified dates and confirms the presence of data for dates where null values were initially suspected. \par
The report's findings are consistent with the statistical characteristics of market data during periods of high volatility and stress. As a statistics expert, I would emphasize the importance of using robust statistical methods to analyze such data, including outlier detection and time-series analysis, to understand the probability and potential impact of similar events in the future.\par
----------------------------------------------------------------------------------------------------------------------------------------------
Moderator:\\
Thank you all for your insightful contributions. It seems that there is a consensus among the management discussants that the expert summary report is accurate and reliable. The report has been thoroughly reviewed from different expert perspectives—financial market, macroeconomic, and statistical—and each of you has confirmed the historical accuracy and statistical significance of the S\&P 500 Index movements on the dates in question.\par
The findings regarding the percentage drop on Black Monday in 1987, the percentage increase during the financial crisis recovery efforts in 2008, and the percentage drop in response to COVID-19 in 2020 are all consistent with historical records and statistical analyses. Additionally, the clarification of the null values for the dates following each event has been acknowledged as crucial for maintaining data integrity and ensuring accurate analysis.\par
Given the agreement on the report's content and the importance of its findings for understanding market dynamics, risk management, policy development, and statistical modeling, it appears that the management team approves the expert summary report as it stands.
Unless there are any further comments or concerns, we can conclude that the expert summary report is accepted and does not require further review. Thank you for your participation and valuable input in this discussion.\par
----------------------------------------------------------------------------------------------------------------------------------------------
Critic:\\
As a critic, I would like to commend the discussants for their thorough review of the expert summary report. The cross-disciplinary perspectives provided by the financial market, macroeconomic, and statistical experts have enriched the analysis and confirmed the accuracy of the historical data and the significance of the market movements on the dates in question.\par
The discussants have effectively contextualized the events of Black Monday, the 2008 financial crisis recovery, and the market's response to the COVID-19 pandemic within their respective fields. The financial market expert highlighted the implications for trading rules and systems, the macroeconomic expert discussed policy interventions and financial stability, and the statistician emphasized the statistical significance and the importance of data integrity for analysis.\par
The clarification regarding the null values for the dates following each event is particularly noteworthy, as it ensures that any analysis based on this data is grounded in complete and accurate information. This is essential for maintaining the integrity of historical market data and for making informed decisions based on that data.\par
The consensus among the discussants suggests that the expert summary report is robust and does not raise any doubts about its findings. The report appears to be a reliable document that can be used for further research, policy-making, and strategic planning.
In conclusion, the discussants have provided a comprehensive review of the expert summary report, and there are no apparent doubts or concerns that need to be addressed. The report is accepted as a valid and valuable resource for understanding the historical performance of the S\&P 500 Index during significant market events.\par
----------------------------------------------------------------------------------------------------------------------------------------------
Moderator:\\
Thank you for your critical assessment and for recognizing the value of the multi-disciplinary review process. It is clear that the expert summary report has been meticulously examined and has stood up to scrutiny from various professional angles. The report's findings have been validated, and the importance of its accuracy and completeness has been emphasized.\par
With the critic's endorsement of the discussants' consensus and the absence of any further doubts or concerns, we can definitively conclude that the expert summary report is approved by the management team. The report will serve as a reliable reference for understanding the historical performance of the S\&P 500 Index during significant market events and will inform future research, policy-making, and strategic decision-making. \par
This concludes our management discussion on the expert summary report. Thank you all for your valuable contributions and for ensuring the report's integrity and utility. The discussion is now officially wrapped up.\par
----------------------------------------------------------------------------------------------------------------------------------------------
Moderator:\\
As the discussion has concluded and the expert summary report has been approved by the management team, there are no further actions required at this moment. If any new issues or topics arise that require attention, please feel free to initiate a new discussion. Thank you, everyone, for your participation.\par
----------------------------------------------------------------------------------------------------------------------------------------------
Critic:\\
As a critic, I have no further feedback to provide. The discussion has been concluded satisfactorily, and the moderator has closed the session. If any new information or topics of interest arise in the future, I look forward to providing critical analysis as needed. Thank you.\par
----------------------------------------------------------------------------------------------------------------------------------------------
Moderator:\\
The discussion is now officially closed. Thank you all for your participation and valuable insights. If any new matters arise that require attention or further discussion, please feel free to initiate a new session. Have a great day!\par
----------------------------------------------------------------------------------------------------------------------------------------------
Critic:\\
As there are no further instructions or content to review, and the discussion has been closed by the moderator, no further action is required at this time. If you have any new queries or topics for discussion, please feel free to initiate a new conversation. Have a great day!\par
\\
\hline
\end{longtable}

\section{Conclusion and Future Direction}
The demonstration of AI in financial market analysis through a multi-agent workflow showcases the potential of emerging technologies to improve data monitoring and anomaly detection. Integrating LLMs with traditional analysis methods could significantly enhance the precision and efficiency of market oversight and decision-making. This approach promises to streamline the review of data, enabling quicker detection of market anomalies and timely information for decision-makers.\par
A central factor in the success and widespread adoption of this approach lies in the effective management of metadata and data governance. Metadata serves as an essential bridge that facilitates the transition of tabular data into a structure conducive to LLM processing, enriching the data's context and enhancing the efficiency and accuracy of the LLM-driven process.\par
This advance in AI-driven financial market data analysis suggests a reconfiguration of the data analysis and decision-making landscape. With ongoing advances in AI technology, the future envisages a framework capable of autonomously executing increasingly complex analytical tasks, diminishing the need for human oversight. This evolution towards an AI-centric approach in financial market data analysis is anticipated not only to streamline anomaly detection and review procedures but also to find applicability in various areas requiring complex data analytical capabilities.\par
Amid these promising developments and the prospective future of AI in financial market analysis, it is crucial to emphasise the indispensable role of human oversight throughout the developmental phases of AI technologies. The imperative for accuracy, accountability, and adherence to ethical standards in AI applications calls for vigilant human supervision. As AI systems gain autonomy and become more integrated in decision-making processes, the potential for systemic biases, inaccuracies, and unintended outcomes underscores the need for ongoing human involvement. Such engagement is essential not only for the validation of AI outputs but also for steering the evolution of these technologies in a direction that aligns with ethical standards and societal values.\par

\bibliographystyle{unsrt}  
\bibliography{references}

\end{document}